\documentclass[aps,pra,twocolumn,nofootinbib,twocolumn,floatfix,showpacs,superscriptaddress]{revtex4-2}
\usepackage{amsmath,graphicx,epsfig,amsthm,color,bm}
\usepackage{pifont,bbding,amssymb,soul,mathrsfs,braket}
\usepackage{dcolumn}
\usepackage{siunitx}
\usepackage[hyperindex,breaklinks,colorlinks=true,citecolor=blue,urlcolor=blue]{hyperref}
\usepackage{subfigure}

\makeatletter

\newcommand{\Rmnum}[1]{\expandafter\@slowromancap\romannumeral #1@}
\makeatother

\begin{document}

\title{Floquet Diamond Sensor with Optimal Precision}
\author{Qi-Tao Duan}
\affiliation{School of Physics, State Key Laboratory of Crystal Materials, Shandong University, Jinan 250100, China}
\author{Teng Li}
\affiliation{School of Physics, State Key Laboratory of Crystal Materials, Shandong University, Jinan 250100, China}
\author{Si-Qi Chen}
\affiliation{School of Physics, State Key Laboratory of Crystal Materials, Shandong University, Jinan 250100, China}
\author{Shengshi Pang}
\affiliation{School of Physics, Sun Yat-sen University, Guangzhou, Guangdong 510275, China}
\affiliation{Hefei National Laboratory, University of Science and Technology of China, Hefei 230088, China}
\author{He Lu}
\email{luhe@sdu.edu.cn}
\affiliation{School of Physics, State Key Laboratory of Crystal Materials, Shandong University, Jinan 250100, China}


\begin{abstract}
The diamond sensor has emerged as a promising platform for quantum sensing, enabling the estimation of physical quantities---such as microwave~(MW) field---with precision unattainable by classical counterpart. However, traditional diamond sensors suffer severe precision degradation when the signal MW is not resonant with the sensor transition frequency. Here, we propose and demonstrate a Floquet diamond sensor~(FDS) for high-precision off-resonant MW amplitude sensing without attenuating the strength of the signal MW. The periodic driven field effectively induces an quasi-energy shift that matches the off-resonant MW frequency. The measurement precision of FDS is characterized by quantum Fisher information, which approaches the ultimate precision---Heisenberg limit---within the coherent time. Furthermore, the FDS exhibits robust tolerance to practical control errors and is compatible with dynamical coupling protocol, enabling a robust and high-sensitivity magnetic sensing. Our results confirm the quantum advantage of quantum sensing and provide a practical technology for high-precision off-resonant MW sensing.
\end{abstract}

\maketitle

Estimation of physical quantities with high precision stands at the core of both science and technology. In recent years, numerous applications have emerged that utilize quantum systems as sensors for physical quantities, wherein quantum features are harnessed to estimate parameters with a precision unattainable by even the most advanced classical strategies~\cite{Giovannetti2004Science,Degen2017RMP,Braun2018RMP}. Notably, nitrogen-vacancy~(NV) centers in diamond constitute an increasingly favored quantum sensing platform~\cite{Jelezko2004PRL,Barry2020RMP}, as the electronic spin defects can be individually addressed, optically polarized and detected, and exhibit excellent coherence properties even at room temperature~\cite{Hanson2006PRL,Childress2006Science}. To date, there is a growing body of research demonstrating diamond sensor for physical quantities including magnetic field~\cite{Aiello2013NC, Boss2017Science, Barry2020RMP, Rovny2022Science}, electric field~\cite{Dolde2011NP, Bian2021NC, Qiu2022NPJQI}, stress~\cite{Hsieh2019Science, Broadway2019NL} and temperature~\cite{Neumann2013NL, Moreva2020PRApplied}. 

Most recently, microwave~(MW) sensing has attracted considerable attention~\cite{Wang2015NC,Horsley2018PRApplied,Chen2023NC}, with potential applications in areas such as wireless communications~\cite{Holl2017PRL}, radar technology~\cite{Barzanjeh2020SA}, nanoscale detection of magnons in spintronic materials~\cite{Lee2020NL} and breast cancer detection~\cite{Wang2018Sensors}. Rabi measurement is a typical sensing protocol, which can provide information not only on the frequency $\omega_s$ but also on the~(transverse) magnitude $\Omega_s$ of signal MW. For the magnitude sensing, the optimal precision necessitates that the signal MW resonates with the transition frequency $\omega_0$ of diamond sensor. However, even a small detuning of the MW can significantly degrade measurement precision, or even render it entirely ineffective~\cite{Carmiggelt2023NC}. While the resonance frequency of diamond sensors can be tuned via a magnetic bias field, precisely adjusting this bias field to match the frequency of the signal MW is time-consuming---thus limiting the real-time sensing applications.  

Several approaches have been developed to detect off-resonant signals without relying on a tunable magnetic bias field. Quantum frequency mixers have been explored, which mix the signal frequency with control frequencies to generate the resonance frequency of diamond sensor~\cite{Wang2022PRX, Karlson2024PRApplied}, thereby enabling broadband MW sensing sensing~\cite{Wang2022PRX, Karlson2024PRApplied}. Hybrid diamond sensor~\cite{Carmiggelt2023NC}, on the other hand, leverages the nonlinearity of the magnet to convert the signal frequency to resonate frequency of diamond sensor. However, both approaches are constrained by limited conversation efficiency, i.e., the magnitude of MW signal is reduces, which in turn degrades the sensing precision.

\begin{figure}[ht!]
    \centering
    \includegraphics[width=\linewidth]{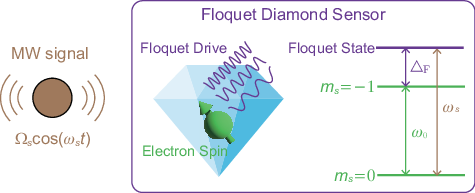}
    \caption{The illustration of Floquet diamond sensor~(FDS). The original diamond sensor~(ODS) is periodically driven by a control field, which introduces a quasi-energy shift that matches the frequency of off-resonant signal MW. }
    \label{Fig:1}
\end{figure}

In this work, inspired by the Floquet engineering~\cite{Goldman2014PRX,Goldman2015PRX,Bukov2015AP,Eckardt2017RMP} and quantum control theory~\cite{Yuan2015PRL,Pang2017NC,Xu2019npjQI,Xu2021PRA,Yang2022PRL}, we propose and demonstrate a periodically driven diamond sensor, so-called Floquet diamond sensor~(FDS), that is able to sense the amplitude of off-resonant MW with optimal precision. As illustrated in Fig.~\ref{Fig:1}, the concept of FDS is to tailor a original diamond sensor~(ODS) by periodic drive, and the Floquet state induced by this drive introduces an quasi-energy shift that matches the frequency of off-resonant signal MW. Notably, the Floquet drive is applied \emph{solely} to the diamond sensor---a design that avoids attenuating the strength of the signal MW. Furthermore, the driving field is engineered using an experimentally feasible control set, which relaxes the stringent requirements for implementing optimal control. The sensing precision of FDS is benchmarked using quantum Fisher information~(QFI)~\cite{Braunstein1994PRL, Braunstein1996AOP, petz2011introduction}, demonstrating that the precision approaches the Heisenberg limit~$\Delta \Omega_s \propto t^{-1}$. Also, we show that the FDS is compatible with dynamical decoupling~(DD)~\cite{Suter2016RMP}, which extends the coherence time of FDS from 17.9~$\mu$s to 162.5~$\mu$s, thereby achieving sensitivity of 195~nT$\cdot$~Hz$^{-1/2}$.

The diamond sensor in our experiment is a single negatively charged NV center, in which the ground state is a triplet manifold of $\ket{m_S=0}$ and $\ket{m_S=\pm1}$ with a zero-field splitting of $D=2 \pi \times 2.87$~GHz. $\ket{m_s=-1}$ and $\ket{m_S=+1}$ are split by applying a static magnetic field $B_0$ along the NV axis.  For ODS, the Hamiltonian within the subspace spanned by $\ket{m_S=0}$ and $\ket{m_S=-1}$~(hereafter referred to as $\ket{0}$ and $\ket{1}$) is given by $\mathcal H_\text{NV} = -\omega_0\sigma_z/2$, where $\omega_0=D-\gamma_eB_0$ denotes the resonance frequency of ODS and $\gamma_e=2 \pi \times 2.8$~MHz/G is the gyromagnetic ratio of the electron. Assume a transverse MW signal with Hamiltonian $\mathcal H_s=\Omega_s\cos(\omega_st)\sigma_x$ couples to the diamond sensor, the Hamiltonian of sensing is given by 
\begin{equation}
\mathcal H_\text{ODS}^\prime=\frac{\Omega_s}{2}\sigma_x+\frac{\omega_s-\omega_0}{2}\sigma_{z}.
\end{equation} 
Hereafter, the notion of Hamiltonians $\mathcal H$, $\mathcal H^\prime$ and $\tilde{\mathcal H}$ denote the forms in the laboratory coordinate system, the rotating frame defined by $U_s=e^{-i\omega_s t\sigma_z/2}$ and the Floquet rotating frame defined bye $U_\text{F} = e^{i K(t)}$, respectively. In Rabi measurement, the ODS is initialized to state $\ket{\psi(0)}=\ket{0}$. Under the evolution governed by $\mathcal H_\text{ODS}^\prime$, the population probabilities of the state $\ket{0}$ evolves as
\begin{equation}
P_0(t)=1-\frac{\Omega_s^2}{\Omega_s^2+\Delta^2}\sin^2\left(\frac{\sqrt{\Omega_s^2+\Delta^2}}{2}t\right).
\end{equation}
where $\Delta=\omega_s-\omega_0$ is the detuning between signal MW and resonance frequency of ODS. The estimation precision of $\Omega_s$ is bounded by the $\rm Cram\acute{e}r$-Rao bound~\cite{Giovannetti2011NP}, which is related to the QFI~\cite{Braunstein1994PRL, Braunstein1996AOP, petz2011introduction}
\begin{equation}
    \mathcal I_{\Omega_s}^{\text{Q}}(t) = 4\left(\langle \partial_{\Omega_s} \psi(t) | \partial_{\Omega_s} \psi(t) \rangle - {| \langle \psi(t) | \partial_{\Omega_s} \psi(t) \rangle |}^2 \right).
    \label{Eq:QFI}
\end{equation}
QFI characterizes the distinguishability of $\ket{\psi(t)}$ with respect to changes in $\Omega_s$. In the resonance case where $\omega_s=\omega_0$, the initial state $\ket{\psi(0)}$ evolves along geodesic of Bloch sphere, and QFI scales as $\mathcal I_{\Omega_s}^{\text{Q}}(t)=t^2$. For $\omega_s\neq\omega_0$, $\mathcal I_{\Omega_s}^{\text{Q}}(t)<t^2$, indicating the optimal precision is bounded by the resonance case. 
\begin{figure*}[t]
    \centering
    \includegraphics[width=\linewidth]{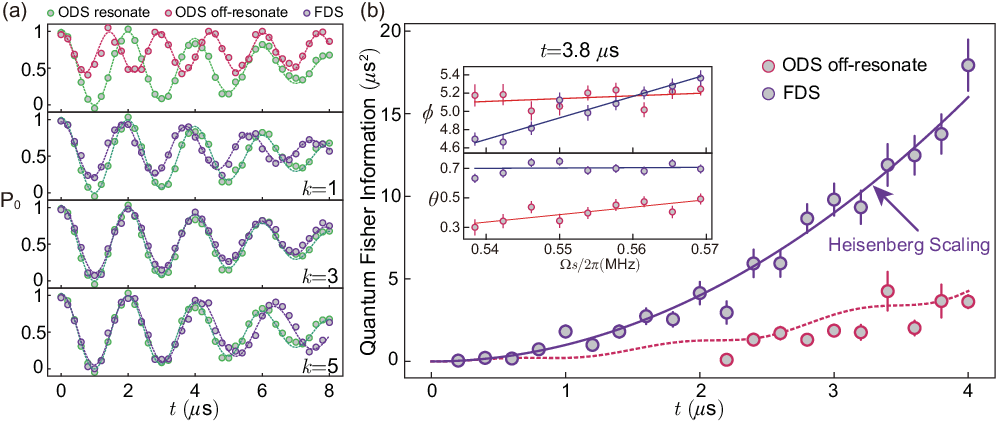}
    \caption{Experimental results of MW sensing. (a) The Rabi oscillations of ODS when sensing the resonant MW signal~(green circles) and the off-resonate MW signal~(red circles). The purple circles are the results of Rabi oscillation of FDS with $k=1, 3$ and 5 when sensing the off-resonant MW signal. (b) The insets shows the $\phi$ and $\theta$ as functions of $\Omega_\text{s}$. Red line is Heisenberg scaling and purple circles are experiment results for QFI of FDS. Red line (circles) is simulation (experiment) results for QFI of ODS. The error bars are calculated by Monte Carlo simulation with Poisson noise.}
    \label{Fig:2}
\end{figure*}

In Floquet diamond sensor (FDS), the diamond sensor is periodically driven by control field of $4\Omega_\text{F} \cos[(\omega_s - \omega_\text{F}) t] \sigma_x$. Consequently, the sensing Hamiltonian of FDS is given by
\begin{equation}
\begin{aligned}
\mathcal H_\text{FDS}^\prime=\frac{\Omega_s}{2}\sigma_x&+\frac{\omega_s-\omega_0}{2}\sigma_{z}\\
&+2\Omega_\text{F} \left[\cos(\omega_\text{F} t) \sigma_x + \sin(\omega_\text{F} t)\sigma_y\right].
\end{aligned}
\end{equation}
The third term is Floquet-driven induced that can be rewritten as $\mathcal H_\text{F}^\prime(t)=\Omega_\text{F}\left(e^{i\omega_\text{F}t}\sigma_-+e^{-i\omega_\text{F}t}\sigma_+\right)$, where $\sigma_\pm=\sigma_x\pm i\sigma_y$ are the raising and lowering operators. $\mathcal H_\text{F}^\prime(t)$ is a periodical Hamiltonian~(i.e., $\mathcal H_\text{F}^\prime(t+T)=\mathcal H_\text{F}^\prime(t)$), so that it is convenient to calculate the dynamics of in the Floquet rotating frame $U_\text{F} = e^{i K(t)}$, where
\begin{equation}
     K(t) = \frac{\Omega_\text{F}}{i\omega_\text{F}}(e^{i \omega_\text{F} t}\sigma_- - e^{-i \omega_\text{F} t}\sigma_+ ) + \mathcal O(\frac{1}{\omega_\text{F}^2})
    \label{Eq:kick}
\end{equation}
is the kick operator~\cite{Goldman2014PRX, Goldman2015PRX}. In this frame, $\mathcal H_\text{F}^\prime(t)$ is written as 
\begin{equation}
      \tilde{\mathcal H}_\text{F} =- \frac{4\Omega_\text{F}^2}{\omega_\text{F}}\sigma_z + \mathcal O(\frac{1}{\omega_\text{F}^2}),  
   \label{Eq:FloquetH}
\end{equation}
which is a time-independent effective Hamiltonian. Accordingly, $\mathcal H_\text{FDS}^\prime$ is
\begin{equation}\label{Eq:FloquetNV}
    \tilde{\mathcal H}_\text{FDS} = \frac{\Omega_s}{2}\sigma_x+\left(\frac{\omega_s-\omega_0}{2}- \frac{4\Omega_\text{F}^2}{\omega_\text{F}}\right) \sigma_z+\mathcal O(\frac{1}{\omega_\text{F}^2}).
\end{equation}
The high-order term $\mathcal O(1/\omega_\text{F}^2)$ can be neglected if $\omega_\text{F}$ is larger enough, i.e., $\omega_\text{F}\gg\Omega_s$, $\Omega_\text{F}$, and $\omega_s-\omega_0$. According to Eq.~\ref{Eq:FloquetNV}, the periodic drive $\mathcal H_\text{F}^\prime(t)$ induces a quasi-energy shift $\Delta_\text{F}=8\Omega_\text{F}^2/\omega_\text{F}$~(also called the AC stark shift), effectively modulating the resonance frequency of diamond sensor to $\omega_0^\text{F}=\omega_0+\Delta_\text{F}$. Note that the Floquet driven doest not physically change the energy levels of diamond sensor. Instead, it causes the evolution of $\ket{0}$ to \emph{appear as} if it occurs in the presence of energy shift. Such a quasi-energy enables the sensing of off-resonant frequency $\omega_s=\omega_0^\text{F}$ with Heisenberg-scaling precision.

Indeed, neglecting the high-order term $\mathcal O(1/\omega_\text{F}^2)$ requires that both $\omega_\text{F}$ and $\Omega_\text{F}$ be sufficiently large, which remains experimentally challenging. In fact, the Floquet diamond sensor can be constructed using driven field with multiple frequencies, thereby alleviating the requirement for a high $\Omega_\text{F}$. Specifically, the Hamiltonian of multi-frequency driven field is in form of 
\begin{equation}
\begin{split}
    \mathcal H_\text{F}^\prime(t) &= 2\Omega_\text{F} \sum_{l=1}^k [\cos(l \omega_\text{F} t) \sigma_x + \sin(l \omega_\text{F} t)\sigma_y]\\
    &= \Omega_\text{F}\sum_{l=1}^k (e^{il \omega_\text{F} t} \sigma_- + e^{-il \omega_\text{F} t} \sigma_+),
    \label{Eq:Hc}
    \end{split}
\end{equation}
with $k$ being a positive integer~\cite{Yang2022PRL}. Similarly, the Floquet rotating frame is defined by $U_\text{F} = e^{iK(t)}$ with $K(t) = \Omega_\text{F}\sum_{l=1}^k(e^{il \omega_\text{F} t} \sigma_- - e^{-il \omega_\text{F} t} \sigma_+)/il\omega_\text{F} + \mathcal O(1/\omega_\text{F}^2)$ with $\mathcal O(1/\omega_\text{F}^2)$ being the high-order terms. By neglecting $\mathcal O(1/\omega_\text{F}^2)$, the Hamiltonian of FDS simplifies to
\begin{equation} \label{Eq:FloquetH}
    \tilde{\mathcal H}_\text{FDS} \approx \frac{\Omega_s}{2}\sigma_x+\left( \frac{\omega_s-\omega_0}{2} - \frac{4}{\omega_\text{F}} \sum_{l}^k \frac{\Omega_\text{F}^2}{l} \right) \sigma_z.
\end{equation}
The Floquet-driven induced energy shift is $\Delta_\text{F}=8 \sum_{l=1}^k \Omega_\text{F}^2/ l\omega_\text{F}$. Compared to single-frequency driving, multi-frequency driven relaxes the requirement on the amplitude $\Omega_\text{F}$.

In our experiment, the resonance frequency of the ODS for the transition $\ket{0}\leftrightarrow\ket{1}$ is $\omega_0=2\pi\times1.47$~GHz under an external magnetic field of $B_0\approx500$~G. Here, $B_0$ is a static magnetic filed aligned along the NV axis~([111] crystallographic direction) using a permanent magnet, which serves to reduce the nuclear spin noise~\cite{Smeltzer2009PRA}. For the signal MW with frequency resonate with ODS~(i.e., $\omega_s=\omega_0$), the measured Rabi oscillation is shown by green circles in Fig.~\ref{Fig:2}~(a), and the strength of $\Omega_s$ can be determined accordingly. As mentioned above, the sensing of $\Omega_s$ is highly sensitive to the resonance. We slightly detune the frequency of signal MW by 0.5~MHz~(i.e., $\omega_s-\omega_0=2\pi\times0.5$~MHz), and the measured Rabi oscillation is shown by red circles in Fig.~\ref{Fig:2}~(a), which clearly results in an imprecise estimation of $\Omega_s$. In FDS, ODS is periodically driven by MW fields $4\Omega_\text{F}  \sum_{l=1}^k \cos[(\omega - l\omega_\text{F}) t] \sigma_x$. We set $\Omega_\text{F}/2\pi=1$~MHz and $\omega_\text{F}=36.54$~MHz, Rabi oscillation results for $k=1, 3$ and $5$ are shown by purple circles in Fig.~\ref{Fig:2}~(a). As $k$ increases, the Rabi oscillation becomes closer to the resonant case, indicating that $\Omega_s$ can be measured with higher precision. 

To further investigate the sensing precision of $\Omega_s$ with FDS, we measure the QFI. In the ideal case, the initial state $\ket{0}$ evolves to $\ket{\psi_{\Omega_s}(t)}=\cos\theta\ket{+}+\sin\theta e^{i\phi}\ket{-}$ with $\ket{\pm}=(\ket{0}\pm\ket{1})/\sqrt{2}$ after a time $t$. According to Eq.~\ref{Eq:QFI}, the QFI for $\Omega_s$ can be rewritten as
\begin{equation}
    \mathcal I_{\Omega_s}^{\text{Q}}(t) = 4\left(\frac{\partial \theta}{\partial \Omega_s}\right)^2 + \sin^2(2\theta)\left(\frac{\partial\phi}{\partial\Omega_s}\right)^2.
    \label{Eq:qfic2}
\end{equation}
To quantify the QFI, we vary $\Omega_s$, measure the expected values $\langle\sigma_x\rangle$, $\langle\sigma_y\rangle$ and $\langle\sigma_z\rangle$ of states $ \ket{\psi_{\Omega_s}(t)}$, and then estimate $\theta$ and $\phi$ by 
\begin{equation}
    \phi = \arctan \left(\frac{-\langle\sigma_y\rangle}{\langle\sigma_z\rangle}\right),
    \theta = \frac{\arccos\langle\sigma_x\rangle}{2}.
    \label{Eq:phi and theta}
\end{equation}
The results of $\phi$ and $\theta$ at evolution time $t=3.8~\mu$s are shown in the insets of Fig.~\ref{Fig:2}~(b). By linear fitting of the data, we obtain the value of $\partial \theta/\partial \Omega_s$ and $\partial\phi/\partial\Omega_s$ and then calculate the value of $\mathcal I_{\Omega_s}^{\text{Q}}$ in Eq.~\ref{Eq:qfic2}. To observe the scaling behavior of the QFI, we vary the evolution time $t$, and the results of QFI~(expressed as $\mathcal I_{\Omega_s}^{\text{Q}}(t)$) are shown in Fig.~\ref{Fig:2}~(b), which agrees well with the Heisenberg scaling~(purple solid line). For comparison, we also measure the QFI of ODS in the off-resonant case, and the results are shown by red circles in Fig.~\ref{Fig:2}~(b), in which the scaling of QFI is far below $t^2$. 

\begin{figure*}[ht]
    \centering
    \includegraphics[width=\linewidth]{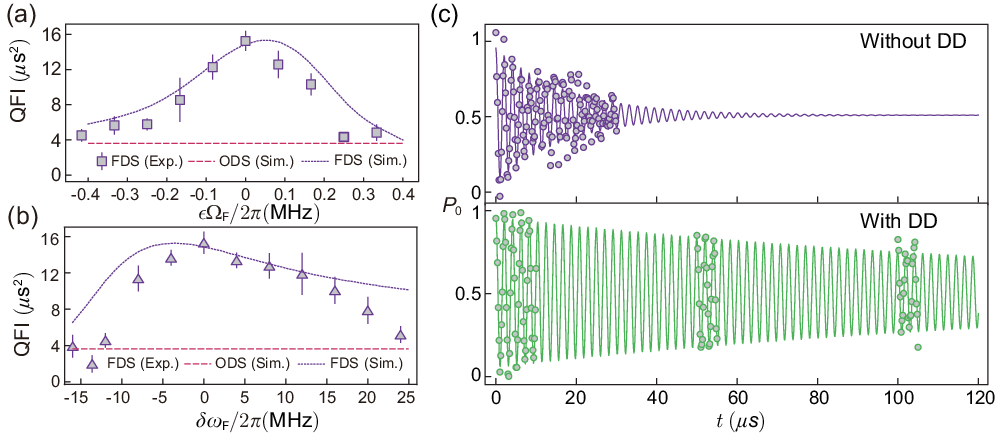}
    \caption{Robustness of Floquet diamond sensor. (a) and (b) respectively illustrate QFI when the parameters $\Omega_\text{F}$ and $\omega_\text{F}$ in the control Hamiltonian do not match the theoretical optimal values. Purple line is numerically simulation results for $\mathcal H_\text{F}^\prime$ and points are experimental results. Red line is QFI of ODS. (c) The Rabi oscillation of the NV center driven by the off-resonant signal for FDS without (top) and with (bottom) DD sequence, the decoherence time is approximately $\SI{17.9}{\mu s}$ and $\SI{162.5}{\mu s}$ respectively.}
    \label{Fig:3}
\end{figure*}

While FDS enables sensing of MW signal with optimal precision, however, Floquet-driven pulses inevitably introduce errors. These pulse errors can generally be categorized into Rabi frequency error and frequency detuning error~\cite{Rong2015NC,Xu2020PRL}. To investigate the robustness of the FDS, we introduce the Rabi error and detuning error in $\mathcal H_\text{F}^\prime$ by setting $\Omega_\text{F}^\prime(t)=\Omega_\text{F}+\epsilon \Omega_\text{F}$ and $\omega_\text{F}^\prime=\omega_\text{F}+\delta\omega_\text{F}$, respectively. The results of QFI with noisy Floquet driven $\mathcal H_\text{F}^\prime(t)$ at $t=4~\mu$s are shown with purple squares and triangles in Fig.~\ref{Fig:3}~(a) and Fig.~\ref{Fig:3}~(b). The QFI of FDS decreases with increasing $|\epsilon\Omega_\text{F}|$ or $|\delta\omega_\text{F}|$. This is because the energy shift $\Delta_\text{F}$ induced by the noisy $\mathcal H_\text{F}^\prime(t)$ does not perfectly compensate for the detuning $\Delta$, thus reducing the precision of FDS. Nevertheless, within the noise ranges of $\epsilon \Omega_\text{F}/2\pi\in[-0.42~\text{MHz}, 0.33~\text{MHz}]$ and $\delta\omega_\text{F}/2\pi\in[-12~\text{MHz}, 24~\text{MHz}]$, the FDS still exhibits enhanced precision compared to ODS~(simulated results shown with red dashed lines in Fig.~\ref{Fig:3}~(a) and Fig.~\ref{Fig:3}~(b)).

The strength of Rabi frequency $\Omega_s$ is proportional to the magnetic field applied in the NV center, following the relation $B_s=\sqrt{2}\Omega_s/\gamma_e$. Consequently, the magnetic sensitivity of FDS also exhibits Heisenberg scaling. Based on Rabi oscillation, the magnetic sensitivity of MW field is given by~\cite{Chen2023NC}
\begin{equation}
    \eta \propto \frac{1}{\gamma_eC\sqrt{N}}\frac{\sqrt{1+t/t_\text{det}}
    }{te^{-t/T_2}} 
    \label{Eq:sensitivity}
\end{equation}
where $C$ denotes fluorescence contrast between $\ket{0}$ and $\ket{1}$, $t_\text{det}$ is the readout time,  $N$ is counting rate of fluorescence and $T_2$ is the coherent time~(i.e., Rabi oscillation decay time). The results of Rabi oscillation of FDS driven by an off-resonant MW signal are shown in Fig.~\ref{Fig:3}~(c). By fitting the decay of oscillation amplitude, we obtain $T_2\approx17.9~\mu$s. In
the experiment, the detection time $t_\text{det}=\SI{0.94}{\mu s}$, contrast $C \approx 0.13$ and fluorescence counting rate $N \approx 9.5 \times 10^4$. The Heisenberg scaling of sensitivity is constrained within the duration of the coherence time $T_2$. According to Eq.~\ref{Eq:sensitivity}, the highest sensitivity 602~nT$\cdot$Hz$^{-1/2}$ would be achieved at $t=T_2$. To extend the coherence time, we implemented a dynamical decoupling (DD) protocol on the FDS, and the results are shown in Fig.~\ref{Fig:3}~(c). With the implement of DD, the coherence time is significantly extended to $T_2\approx162.5~\mu$s, yielding a highest sensitivity of  195~nT$\cdot$Hz$^{-1/2}$. More details can be found in Supplementary Materials.

In conclusion, we propose and demonstrate a FDS to address the critical challenge of degraded precision in off-resonant MW sensing with diamond senor. Experimental results convincingly show the improvement of measurement precision of FDS, i.e., the results of QFI demonstrate that FDS achieves Heisenberg limit precision within the coherence time. Additionally, the FDS shows robust tolerance to practical control errors, i.e., Rabi frequency and frequency detuning errors, maintaining enhanced QFI over a broad range of parameter deviations. The performance of FDS is further improved with DD protocol, which extends the coherent time about one order of magnitude and thus achieves highest magnetic sensitivity of 195~nT$\cdot$Hz$^{-1/2}$. Our work not only provides a practical technology for high-precision off-resonant MW sensing but also empowers the application of Floquet engineering in quantum sensing. The compatibility of the FDS with existing NV center sensing schemes and its robustness to control errors make it promising for real-world applications.
 
\appendix
\section{MW sensing with Rabi oscillation}
The electron spin in diamond sensor is an two-level quantum system, i.e., $\ket{0,+1}=\ket{0}$ and $\ket{-1,+1}=\ket{1}$. The corresponding Hamiltonian~($\hbar=1$) is $\mathcal H_\text{NV}=-\omega_0\sigma_z/2$. The signal MW field interacting with the diamond sensor is a monochromatic field with frequency $\omega_s$ 
\begin{equation}
S(t)=A\cos(\omega_st).
\end{equation}
Here, the MW field is polarized in $x$-direction perpendicular to the NV axis. The wavelength of $S(t)$ is much larger than the size of single NV center so that the spatial dependence of $S(t)$ can be ignored, which is so-called dipole approximation. In the dipole approximation, the interaction Hamiltonian can be written as 
\begin{equation}
\begin{split}
\mathcal H_I&=\sigma_xS(t)\\
&=A\sigma_x\cos(\omega_st)\\
&=\frac{\Omega_s}{2}\sigma_x\left(\text{e}^{-i\omega_st}+\text{e}^{i\omega_st}\right),
\end{split}
\end{equation}
where $\Omega_s=A=\gamma_e B_s/\sqrt{2}$ is the Rabi frequency. The sensing process can be described by the Hamiltonian
\begin{equation}
\begin{split}
\mathcal H_\text{ODS}&=\mathcal H_\text{NV}+\mathcal H_I\\
&=-\frac{\omega_0}{2}\sigma_z+\frac{\Omega_s}{2}\sigma_x\left(\text{e}^{-i\omega_st}+\text{e}^{i\omega_st}\right).
\end{split}
\end{equation}
The calculations can be simplified by moving to a rotating frame rotating at the driving frequency $\omega_s$. To this end, a unitary rotation operator $U_s(t)=\text{e}^{-i\omega_st\sigma_z/2}$ is introduced, and the Schr\"odinger equation is rewritten in the frame defined by $U_s(t)$
\begin{equation}
i\frac{\partial}{\partial t}\ket{\psi^\prime(t)}=\mathcal H_\text{ODS}^\prime\ket{\psi^\prime(t)},
\end{equation}
where $\ket{\psi^\prime(t)}=U_s(t)\ket{\psi(t)}$ and $\mathcal H_\text{ODS}^\prime=U_s(t) \mathcal H_\text{ODS} U_s(t)^\dagger+i\frac{\partial U_s(t)}{\partial t}U_s(t)^\dagger$. Note that $i\frac{\partial U_s(t)}{\partial t}U_s(t)^\dagger=\frac{\omega_s}{2}\sigma_z$. The first term in $\mathcal H_\text{ODS}^\prime$ can be expanded
\begin{equation}
\begin{split}
U_s(t) \mathcal H_\text{ODS} U_s(t)^\dagger&=U_s(t)\left(\mathcal H_\text{NV}+\mathcal H_I\right)U_s(t)^\dagger\\
&=\mathcal H_\text{NV}+U_s(t)\mathcal H_IU_s(t)^\dagger\\
&=-\frac{\omega_0}{2}\sigma_z+\text{e}^{-i\frac{\omega_st}{2}\sigma_z}\sigma_x\Omega_s\cos(\omega_st)\text{e}^{i\frac{\omega_st}{2}\sigma_z}.
\end{split}
\end{equation}
Using Baker-Campbell-Hausdorff lemma, the second term is 
\begin{widetext}
\begin{equation}
\begin{split}
&\Omega_s\cos(\omega_st)\text{e}^{-i\omega_st\sigma_z/2}\sigma_x\text{e}^{i\omega_st\sigma_z/2}\\
&=\Omega_s\cos(\omega_st)\left\{\sigma_x+\left(\frac{-i\omega_st}{2}\right)[\sigma_z, \sigma_x]+\left(\frac{-i\omega_st}{2}\right)^2\frac{1}{2!}\left[\sigma_z, \left[\sigma_z, \sigma_x\right]\right]+\cdots \right\}\\
&=\Omega_s\cos(\omega_st)\left[\sigma_x\cos(\omega_st)+\sigma_y\sin(\omega_st)\right]\\
&=\frac{\Omega_s}{2}[\sigma_x(1+\cos2\omega_st)+\sigma_y\sin2\omega_st].
\end{split}
\end{equation}
\end{widetext}
As we are focusing on the slow dynamics of the Hamiltonian, we can make the rotating-wave approximation~(RWA) to get rid of the rapidly oscillating terms $\cos(2\omega_st)$ and $\sin(2\omega_st)$. Consequently, the full Hamiltonian in the rotating frame to be written as
\begin{equation}
\mathcal H_\text{ODS}^\prime=-\frac{\omega_0}{2}\sigma_z+\frac{\Omega_s}{2}\sigma_x+\frac{\omega_s}{2}\sigma_z=\frac{\Delta}{2}\sigma_z+\frac{\Omega_s}{2}\sigma_x,
\end{equation}
where $\Delta=\omega_s-\omega_0$ represents the detuning. 

Assuming the ODS is initialized into $\ket{0}$, in the case that the dampings are negligible, the population probabilities of the state $\ket{0}$ evolves as
\begin{equation}
P_0(t)=1-\frac{\Omega_s^2}{\Omega_s^2+\Delta^2}\sin^2\left(\frac{\sqrt{\Omega_s^2+\Delta^2}}{2}t\right).
\end{equation}
The oscillatory behavior of population inversion is well known as Rabi oscillation, and the information of $\Omega_s$ and $\Delta$ can be extracted from the Rabi oscillation.

\section{Floquet diamond sensor}
In Floquet diamond sensor~(FDS), the NV center is driven by multi-frequency fields \begin{equation}
F(t)=F\sum_{l=1}^k\left[\cos(\omega_s-l\omega_\text{F}) t\right],
\end{equation}
with Hamiltonian of 
\begin{equation}
\begin{split}
\mathcal H_\text{NV}^\text{F}(t)&=\mathcal H_\text{NV}+\mathcal H_\text{F}(t)\\
&=-\frac{\omega_0}{2}\sigma_z+4\Omega_\text{F}\sum_{l=1}^k\left[\cos(\omega_s-l\omega_\text{F}) t\right]\sigma_x,
\end{split}
\end{equation}
where $4\Omega_\text{F}=F$ is the Rabi frequency. Interacting with signal MW field, the full Hamiltonian is 
\begin{equation}\label{Eq:HFDSlab}
\begin{split}
\mathcal H_\text{FDS}(t)&=\mathcal H_\text{NV}+\mathcal H_\text{F}(t)+\mathcal H_I\\
&=-\frac{\omega_0}{2}\sigma_z+4\Omega_\text{F}\sum_{l=1}^k\left[\cos(\omega_s-l\omega_\text{F}) t\right]\sigma_x.
\end{split}
\end{equation}

In rotation frame of $U_s(t)$, Eq.~(\ref{Eq:HFDSlab}) is converted to  
\begin{equation}
\begin{aligned}
\mathcal H_\text{FDS}^\prime(t)&=U_s(t)\mathcal H_\text{FDS}U_s^\dagger(t)+i\frac{\partial U_s(t)}{\partial t}U_s^\dagger(t)\\
&=\frac{\Delta}{2}\sigma_z+\frac{\Omega_s}{2}\sigma_x+2\Omega_\text{F} \sum_{l=1}^k [\cos(l \omega_\text{F} t) \sigma_x + \sin(l \omega_\text{F} t)\sigma_y]\\
&=\frac{\Delta}{2}\sigma_z+\frac{\Omega_s}{2}\sigma_x+ \Omega_\text{F}\sum_{l=1}^k (e^{il \omega_\text{F} t} \sigma_- + e^{-il \omega_\text{F} t} \sigma_+).\\
\end{aligned}
\end{equation}
Note that $\mathcal H_0=\frac{\Delta}{2}\sigma_z+\frac{\Omega_s}{2}\sigma_x$ is time-independent, while $\mathcal H^\prime_\text{F}(t)=\Omega_\text{F}\sum_{l=1}^k (e^{il \omega_\text{F} t} \sigma_- + e^{-il \omega_\text{F} t} \sigma_+)$ is periodical, i.e., $\mathcal H_\text{F}^\prime(t+T)=\mathcal H_\text{F}^\prime(t)$. To obtain the time-independent effective Hamiltonian, we consider the unitary transformation 
\begin{equation}
\ket{\tilde\psi(t)}=U_\text{F}(t)\ket{\psi^\prime(t)}=e^{iK(t)}\ket{\psi^\prime(t)},
\end{equation}
where $\ket{\psi^\prime(t)}$ is the solution of Schr\"odinger equation
\begin{equation}
i\frac{\partial}{\partial t}\ket{\psi^\prime(t)}=\mathcal H_\text{FDS}^\prime(t)\ket{\psi^\prime(t)}.
\end{equation}
Then, the Schr\"odinger equation is rewritten as
\begin{equation}
i\frac{\partial}{\partial t}\ket{\tilde\psi(t)}=\Tilde{\mathcal H}_\text{FDS}\ket{\tilde\psi(t)},
\end{equation}
with 
\begin{equation}
    \Tilde{\mathcal H}_\text{FDS} = U_\text{F} \mathcal H_\text{FDS}^\prime(t) U_\text{F}^\dagger +i\frac{\partial U_\text{F}}{\partial t}U_\text{F}^\dagger,
    \label{Eq:effect transform}
\end{equation}
being a time-independent effective Hamiltonian. The analytical expression of $K(t)$ and $\mathcal H_\text{eff}$ is generally challenging, and it is convenient to construct these operators perturbatively by expanding them in the powers of the period $T=2\pi/\omega_\text{F}$~\cite{Goldman2014PRX}. Following Ref.~\cite{Rahav2003PRA,Goldman2014PRX}, by writing 
\begin{equation}
\Tilde{\mathcal H}_\text{FDS} = \sum_{n=0}^k \frac{1}{\omega^n} \mathcal H_\text{eff}^{(n)},\\
K(t)=\sum_{n=1}^k \frac{1}{\omega^n} K^{(n)}(t),\label{Eq:highorder}
\end{equation}
and taking the expansion of Eq.~(\ref{Eq:effect transform})
\begin{widetext}  
\begin{equation}\label{Eq:transform1}
U_\text{F} \mathcal H_\text{FDS}^\prime(t) U_\text{F}^\dagger = \mathcal H_\text{FDS}^\prime(t) + i[K(t), \mathcal H_\text{FDS}^\prime(t)]-\frac{1}{2}[K(t), [K(t), \mathcal H_\text{FDS}^\prime(t)]]-\frac{i}{6}[K(t),[K(t), [K(t), \mathcal H_\text{FDS}^\prime(t)]]] \cdots, 
\end{equation}
\begin{equation}
\frac{\partial U_\text{F}}{\partial t}U_\text{F}^\dagger=i\frac{\partial K(t)}{\partial t}-\frac{1}{2}[K(t), \frac{\partial K(t)}{\partial t}]-\frac{i}{6}[K(t), [K(t), \frac{\partial K(t)}{\partial t}]] \cdots, 
    \label{Eq:transform2}
\end{equation}
\end{widetext}  
we can determine $\Tilde{\mathcal H}_\text{FDS}$ and $K(t)$ at the desired order $\mathcal O(1/\omega_\text{F}^n)$. Note that $\omega_\text{F}$ is assumed to be large enough, resulting $T$ to be small in the expansion procedure. Specifically, $K(t)$ can be chosen as 
\begin{widetext}  
\begin{equation}
    \begin{split}
        K(t)= &\frac{1}{i \omega_\text{F}}\sum_{l=1}^k \frac{1}{l}(\Omega_\text{F}\sigma_-e^{il\omega_\text{F} t} - \Omega_\text{F}\sigma_+e^{-il\omega_\text{F} t})+\frac{1}{i\omega_\text{F}^2} \sum_{l=1}^k \frac{1}{l^2}([\Omega_\text{F}\sigma_-, \mathcal H_0]e^{il\omega_\text{F} t}-\text{H.c.}) \\
        &+ \frac{1}{2i\omega_\text{F}^2}\sum_{l,m=1}^k \frac{1}{l(l+m)}([\Omega_\text{F}\sigma_-, \Omega_\text{F}\sigma_-]e^{i(l+m)\omega_\text{F} t}-\text{H.c.}) + \frac{1}{2i\omega_\text{F}^2}\sum_{l \neq m=1}^k \frac{1}{l(l-m)}([\Omega_\text{F}\sigma_-, \Omega_\text{F}\sigma_+]e^{i(l-m)\omega_\text{F} t}-\text{H.c.})+\cdots\\
        &= \frac{1}{i \omega_\text{F}}\sum_{l=1}^k \frac{1}{l}(\Omega_\text{F}\sigma_-e^{il\omega_\text{F} t} - \Omega_\text{F}\sigma_+e^{-il\omega_\text{F} t})+\mathcal O(\frac{1}{\omega_\text{F}^2}).
    \end{split}
    \label{Eq:K expansion}
\end{equation}
\end{widetext}  
Then, the $\Tilde{\mathcal H}_\text{FDS}$ is obtained according to Eqs.~(\ref{Eq:effect transform}), (\ref{Eq:transform1}) and (\ref{Eq:transform2})
\begin{equation}
    \begin{split}
        \Tilde{\mathcal H}_\text{FDS} = & \mathcal H_0 + \frac{1}{\omega_\text{F}} \sum_{l=1}^k \frac{\Omega_\text{F}^2}{l}[\sigma_-, \sigma_+] +\mathcal O(\frac{1}{\omega^2})\\
        =&\frac{\Delta}{2}\sigma_z-\frac{4\Omega_\text{F}^2}{\omega_\text{F}} \sum_{l=1}^k \frac{1}{l}\sigma_z +\mathcal O(\frac{1}{\omega^2}).
    \end{split}
    \label{Eq:H expansion}
\end{equation}
If $\omega_\text{F}$ is sufficiently large, i.e., $\omega_\text{F} \gg \Omega_\text{F}$, $\Omega_s$, $\Delta$, the high order term $\mathcal O(1/\omega_\text{F}^2) \approx0$ can be neglected. 
\begin{equation}
\begin{split}
\Tilde{\mathcal H}_\text{FDS}=\left(\frac{\Delta}{2}-\frac{4\Omega_\text{F}^2}{\omega_\text{F}} \sum_{l=1}^k \frac{1}{l}\right)\sigma_z+\frac{\Omega_s}{2}\sigma_x.
\end{split}
\end{equation} 
The periodical driven field induces an quasi-energy shift $\Delta_\text{F}= 8 \sum_l^k\Omega_\text{F}^2/l\omega_\text{F}$ in Floquet rotating frame $U_\text{F}$. 

\section{Quantum Fisher Information for estimating $\Omega_s$}
In MW sensing with diamond sensor, the initial state $\ket{0}$ evolves to $\ket{\psi(t)} = \cos\theta\ket{+}+ \sin\theta e^{i\phi}\ket{-}$ after evolution time $t$, where $\theta$ and $\phi$ are all determined by $ \Omega_s $ and $t$. According to the definition of QFI  \cite{Braunstein1994PRL, Braunstein1996AOP, petz2011introduction}
\begin{equation}
    \mathcal I_{\Omega_s}^{\text{Q}}(t) = 4(\langle \partial_{\Omega_s} \psi(t)|\partial_{\Omega_s} \psi(t) \rangle - |\langle \psi(t)|\partial_{\Omega_s} \psi(t) \rangle|^2),
    \label{Eq:QFI formula}
\end{equation}
the first and second terms are 
\begin{widetext}
\begin{equation}\label{Eq:QFI1term}
    \begin{aligned}
        \langle \partial_{\Omega_s} \psi(t)|\partial_{\Omega_s} \psi(t) \rangle &= \left[-\sin\theta\frac{\partial \theta}{\partial \Omega_s} \bra{+} + \left(\cos\theta\frac{\partial \theta}{\partial \Omega_s}-i\sin\theta\frac{\partial\phi}{\partial\Omega_s}\right)e^{-i\phi}\bra{-}\right] \\
        &\times\left[-\sin\theta\frac{\partial \theta}{\partial \Omega_s} \ket{+} + \left(\cos\theta\frac{\partial \theta}{\partial \Omega_s}+i\sin\theta\frac{\partial\phi}{\partial\Omega_s}\right)e^{i\phi}\ket{-}\right] \\
        &= \sin^2\theta\left(\frac{\partial \theta}{\partial \Omega_s}\right)^2 + \cos^2\theta\left(\frac{\partial \theta}{\partial \Omega_s}\right)^2 + \sin^2\theta\left(\frac{\partial \phi}{\partial \Omega_s}\right)^2 \\
        &=\left(\frac{\partial \theta}{\partial \Omega_s}\right)^2 + \sin^2\theta\left(\frac{\partial \phi}{\partial \Omega_s}\right)^2,
    \end{aligned}
\end{equation}
\begin{equation}\label{Eq:QFI2term}
    \begin{split}
        |\langle \psi(t)|\partial_{\Omega_s} \psi(t) \rangle|^2 &= \left|\left(\cos\theta \bra{+} + \sin\theta e^{-i\phi}\bra{-}\right)\times\left[-\sin\theta\frac{\partial \theta}{\partial \Omega_s} \ket{+} + \left(\cos(\theta)\frac{\partial \theta}{\partial  \Omega_s}+i\sin\theta\frac{\partial\phi}{\partial\Omega_s}\right)e^{i\phi}\ket{-}\right]\right|^2 \\
        &= \left|-\sin\theta\cos\theta\frac{\partial \theta}{\partial \Omega_s} + \sin\theta\cos\theta\frac{\partial \theta}{\partial \Omega_s}+i\sin^2\theta\frac{\partial \phi}{\partial \Omega_s}\right|^2 \\
        &= \left|i\sin^2(\theta)\frac{\partial \phi}{\partial \Omega_s}\right|^2 \\
        &= \sin^4\theta\left(\frac{\partial \phi}{\partial \Omega_s}\right)^2.
    \end{split}
\end{equation}
\end{widetext}
By substituting Eqs.~(\ref{Eq:QFI1term}) and (\ref{Eq:QFI2term}) back to Eq.~(\ref{Eq:QFI formula}), the QFI of $\Omega_s$ is 
\begin{equation} \label{Eq:qfii}
    \begin{split}
        \mathcal I_{\Omega_s}^{\text{Q}} &= 4\left[\left(\frac{\partial \theta}{\partial \Omega_s}\right)^2 + \sin^2\theta\left(\frac{\partial \phi}{\partial \Omega_s}\right)^2 - \sin^4\theta{\left(\frac{\partial \phi}{\partial \Omega_s}\right)}^2\right] \\
        &=4\left(\frac{\partial \theta}{\partial \Omega_s}\right)^2 + 4\sin^2\theta\left(1-\sin^2\theta\right)\left(\frac{\partial\phi}{\partial\Omega_s}\right)^2\\
        &=4\left(\frac{\partial \theta}{\partial \Omega_s}\right)^2 + 4\sin^2\theta\cos^2\theta\left(\frac{\partial\phi}{\partial\Omega_s}\right)^2\\
        &=4\left(\frac{\partial \theta}{\partial \Omega_s}\right)^2 + \sin^2(2\theta)\left(\frac{\partial\phi}{\partial\Omega_s}\right)^2.
    \end{split}
\end{equation}
In the resonate case, i.e., $\Delta=\omega_s-\omega_0=0$, the initial state $\ket{0}$ evolves as $\ket{\psi(t)}=\left(\ket{+}+e^{i\Omega_st}\ket{-}\right)/\sqrt{2}$. This evolution corresponds to $\theta=\pi/4$ and $\phi=\Omega_st$, which yields the optimal precision $\mathcal I_{\Omega_s}^{\text{Q}}(t)=t^2$. 

\section{Experimental details}
\subsection{Atomic and energy-level structure of diamond sensor}
\begin{figure}[ht]
    \centering
    \includegraphics[width=\linewidth]{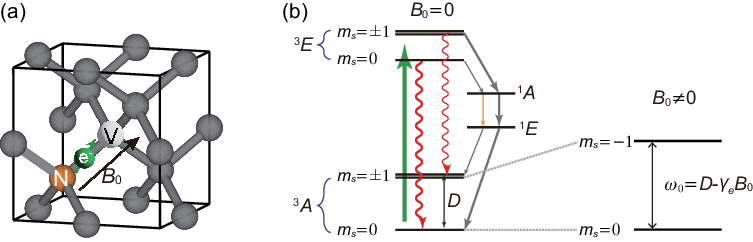}
    \caption{(a) The atomic structure of the nitrogen vacancy~(NV) center
in diamond lattice. (b) Energy-level configuration of the NV defect center.}
    \label{Fig:S1}
\end{figure}
\begin{figure*}[ht!]
    \centering
    \includegraphics[width=\linewidth]{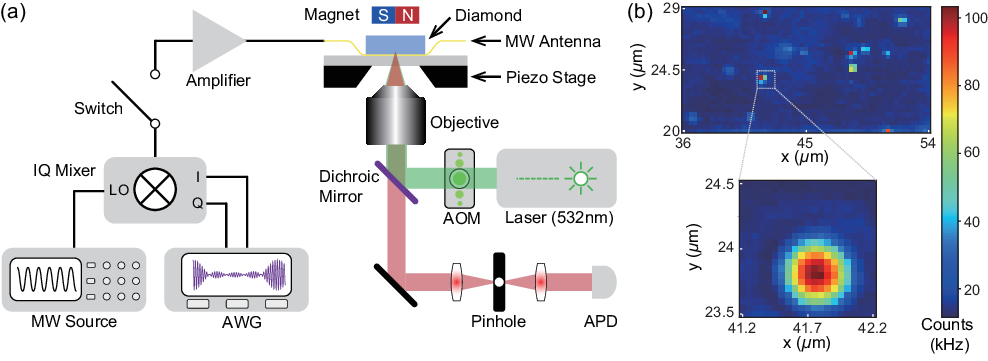}
    \caption{(a) Schematic drawing of the home-built optical confocal microscope setup. (b) The fluorescence intensity map of a $18~\mu\text{m}\times 9~\mu\text{m}$ from a two-dimensional laterally scanning.}
    \label{Fig:setup}
\end{figure*}
The atomic structure of a single nitrogen vacancy~(NV) center in diamond lattice is shown in Fig.~\ref{Fig:S1}~(a), where the gray, orange and white spheres represent the carbon~(C) atom, nitrogen~(N) atom and vacancy site respectively. Notably, the NV center can trap an extra electron~(greed sphere), thereby forming the negatively charged NV$^-$ center. Hereafter, the NV$^-$ center is referred to as NV center for simplicity. The electrons bound to NV center have spin $S=1$, forming a triplet manifold $^3$A and $^3$E ground and excited states as shown in Fig.~\ref{Fig:S1}~(b), respectively. In the absence of an external magnetic field~($B_0=0$), the two $m_s=\pm1$ states are degenerate. In the $^3$A ground states,  the zero-field splitting between $m_s=0$ and $m_s=\pm1$ is approximately $D\approx2.87$~GHz. The degeneracy of the $m_s=\pm1$ states can be lifted by the Zeeman effect via application of a magnetic field along the symmetry axis of the NV center. The original diamond sensor~(ODS) in this experiment is carried out based on the spin transitions between $m_s=0$ and $m_s=-1$ substates with frequency of $\omega_0=D-\gamma_eB_0$.

\subsection{Experimental setup} 
We use a home-built confocal microscope for the selective optical excitation and detection of fluorescence from single NV centers~\cite{Chen2025PRApplied}, as illustrated in Fig.~\ref{Fig:setup}~(a). The excitation light~(532~nm) from a diode laser is digitally modulated by an acousto-optic modulator~(AOM) and then reflected to an oil-immersion objective lens~($\text{NA}=1.25$) by a dichroic mirror. The fluorescence from NV center is collected by the same microscope objective, which transmits the DM and is focused to achieve spatial filtering by a pinhole. The fluorescence passes the pinhole and is then collimated by a second lens and subsequently sent to the  avalanche photodiode~(APD) for detection. 

The diamond sample hosting NV centers is glued on top of a tapered coplanar waveguide~(CPW) board using UV glue with low fluorescence. The CPW is fabricated on the top of a dielectric substrate of quartz glass. Due to the thin copper film and transparency of the glass, the oil-objective lens can be positioned close to the diamond, which benefits the excitation and fluorescence collection. A holder printed circuit board~(PCB) is designed to connect CPW board and MW source via subminiature adapters~(SMAs). Microwave signals are generated by modulating the local oscillator signals with in-phase~(I) and quadrature~(Q) components from an arbitrary waveform generator~(AWG), and subsequently delivered to the NV center via a coplanar waveguide microwave antenna to drive electron spin. The whole sample holder is attached on a $xy$ piezo stage, which can be scanned by $70~\mu\text{m}\times 70~\mu\text{m}$ to locate a single NV center. Fig.~\ref{Fig:setup}~(b) is the fluorescence intensity map of a $18~\mu\text{m}\times 9~\mu\text{m}$ from a two-dimensional laterally scanning. 

\subsection{Spin initialization and readout}
As shown in Fig.~\ref{Fig:S1}~(b), the $^3$A ground states can be optically excited to the $^3$E excited state via spin conserving transitions ($\Delta m_s=0$) by a 532~nm laser. Following excitation, optical relaxation occurs through two pathways: either radiative transitions~($\Delta m_s=0$), which produce broadband red photoluminescence~(PL), or or non-radiative intersystem crossing~(ISC) to the metastable singlet states $^1$E and $^1$A. Notably, non-radiative ISC exhibits strong spin selectivity---specifically, the probability of non-radiative ISC from $m_s=0$ is much smaller than that from $m_s=\pm1$~\cite{Manson2006PRB}.  This spin-selectivity of the decay process enables the polarization of electron spin states into $m_s=0$ after a few optical pumping cycles~\cite{Robledo2011NJP}. This also enables the readout of spin states $m_s=0$ and $m_s=\pm1$ as the spin state $m_s=0$ is brighter than $m_s=\pm1$.
\begin{figure}[ht]
    \centering
    \includegraphics[width=\linewidth]{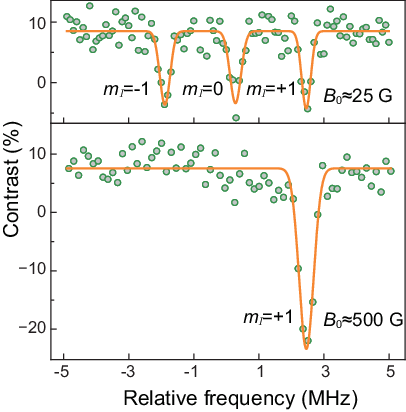}
    \caption{ODMR spectrum over the $m_s=0$ to $m_s=-1$, showing the $^{14}$N hyperfine structure. At low field $B_0\approx25$~G, the spin population is evenly distributed and three lines are visible. At $B_0\approx500$~G, the nuclear spin is polarized into $m_I =+1$.}
    \label{Fig:S3}
\end{figure} 

The optical pumping of the NV center can under certain circumstances lead to nuclear spin polarization~\cite{Jacques2009PRL,Steiner2010PRB}. In our experiment, the nitrogen atom associated to NV center is a $^{14}$N isotope with spin $I=1$, each electron spin state is further split into three hyperfine substates. Spin states will be denoted by $\ket{m_s, m_I}$ in the following. At a magnetic field of $B_0\approx$500~G, level anti-crossing~(LAC) occurs between sublevels $\ket{0, 0}$ and $\ket{-1,+1}$, respectively, $\ket{0, -1}$ and $\ket{-1,0}$ of $^3$E, enabling energy-conserving flip-flop processes between electron and nuclear spin. The spin mixing is not possible for spin states $\ket{0, +1}$ and $\ket{-1, -1}$, so that the spin state is polarized on $\ket{0, +1}$ after the optical pumping. The frequency between $m_s=0$ and $m_s=-1$ is determine by the optically detected magnetic resonance~(ODMR) spectroscopy. As shown in Fig.~\ref{Fig:S3}, At low field $B_0\approx25$~G, the spin population can be seen to be evenly distributed between the three hyperfine states, i.e., $m_I=0, \pm1$. At $B\approx 500$~G, the nuclear spin is polarized into $m_I=+1$.  Experimentally, the strength of $B_0$ is carefully aligned along the NV axis by a permanent magnet held on a three-axis transition stage. The nuclear spin polarization is verified by the vanishing lines of $m_I=-1$ and $m_I=0$.

\subsection{MW sensing sequence}

\begin{figure}[ht]
    \centering
    \includegraphics[width=\linewidth]{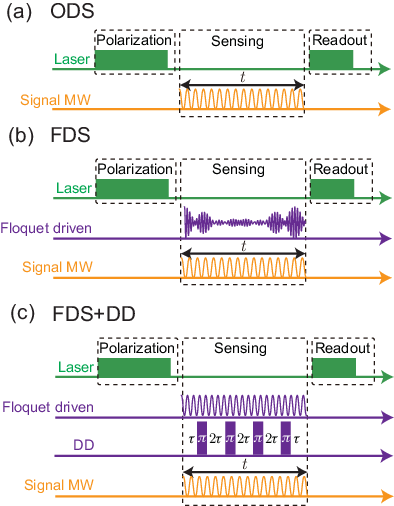}
    \caption{The sensing sequence with (a) ODS, (b) FDS and (c) FDS+DD, respectively.}
    \label{Fig:S4}
\end{figure}

The typical manipulation sequence used to observe Rabi oscillations with ODS is illustrated in Fig.~\ref{Fig:S4}~(a). In the experiment, the ODS is first optically spin-polarized into $m_S=0$ using a 532~nm laser pulse with duration of $5~\mu$s. Subsequently, a waiting interval~(typically approximately $0.3~\mu$s) is introduced to ensure that the ODS can relax from the metastable singlet states to the desired $m_S=0$. The ODS is interacted with a MW signal for time $t$, and then detect using a second 532~nm laser pulse with duration of $t_\text{det}=0.94~\mu$s. In FDS, the Floquet driving is applied during the sensing as shown in Fig.~\ref{Fig:S4}~(b). The decoherence time of FDS can be further extended by applying the CP sequence during the sensing~\cite{Carr1954PR, Suter2016RMP}, as shown in Fig.~\ref{Fig:S4}~(c). In CP sequence, $\pi$-pulse rotates the state around by $\pi$ and $2\tau$ is the delay between the two $\pi$-pulses. In our experiment, we set $\tau=0.5~\mu$s.

\bibliography{reference}

\end{document}